\documentclass[%
    superscriptaddress,
    preprint,
    nofootinbib,
    amsmath,amssymb,
    aps,
    physrev,
]{revtex4-2}

\usepackage[dvipdfmx]{graphicx}
\usepackage[svgnames]{xcolor}
\usepackage{dcolumn}
\usepackage{bm}
\usepackage[utf8]{inputenc}
\usepackage{physics}
\usepackage[
     colorlinks        = true,
     unicode           = true,
     pdfstartview      = FitV,
     linktocpage       = true,
     linkcolor         = OrangeRed,
     citecolor         = MediumSeaGreen,
     urlcolor          = RoyalBlue,
     bookmarks         = true,
     bookmarksnumbered = true,
     breaklinks=true,
     pdftitle={Bias-preserving computation with the bit-flip code},
     pdfauthor={Shoichiro Tsutsui and Keita Kanno}
]{hyperref}
\usepackage[version=3]{mhchem}
\usepackage{siunitx}
\usepackage{qcircuit}
\usepackage{mhchem}

\usepackage{subcaption}

\newcommand{\calE}{\mathcal{E}}

\newcommand{\ini}{\lstick{\ket{0}}}
\newcommand{\iniX}{\lstick{\ket{+}}}

\newcommand{\CZ}{\mathrm{CZ}}

\begin{document}

\title{Bias-preserving computation with the bit-flip code}

\newcommand{\qunasys}{QunaSys Inc., Aqua Hakusan Building 9F, 1-13-7 Hakusan, Bunkyo, Tokyo 113-0001, Japan}
\author{Shoichiro Tsutsui}
\email{tsutsui@qunasys.com}
\affiliation{\qunasys}

\author{Keita Kanno}
\email{kanno@qunasys.com}
\affiliation{\qunasys}

\date{\today}

\begin{abstract}
We explore the feasibility of fault-tolerant quantum computation using the bit-flip repetition code in a biased noise channel where only the bit-flip error can occur. 
While several logic gates can potentially produce phase-flip errors even in such a channel, we propose bias-preserving implementation of $S$, $H$, CZ, and $R_z$ gates. We demonstrate that our scheme improves the computational precision in several tasks such as the time evolution of quantum systems and variational quantum eigensolver.
\end{abstract}

\maketitle

\section{Introduction}
Quantum error correction plays a pivotal role in successfully executing quantum algorithms on noisy quantum computers. Specifically, the surface code~\cite{Kitaev1997, Kitaev_2003, bravyi1998quantum} and its variants exhibit promising attributes concerning scalability. The family of quantum low-density parity check codes~\cite{gottesman2014overhead} that outperform those codes are also actively being investigated. While these error-correcting codes hold the potential to be deployed in fault-tolerant quantum computers in the future, their current utilization in practical calculations is constrained by existing technical limitations.

A complete error-correcting code can correct both bit-flip and phase-flip errors. In actual quantum computers, however, these errors do not appear with equal probability and are generally biased~\cite{GourlayPhysRevA.62.022308,Ioffe_2007,Stephens_2008,Webster_2015,Robertson_2017,fellousasiani2023scalable}. 
For example, phase-flip errors dominate noisy quantum computers employing superconducting qubits~\cite{Aliferis_2009} and Rydberg atoms~\cite{cong2022hardwareefficient}. The same holds true for cat qubits employed in optical quantum computers~\cite{Guillaud_2019,Puri_2020,Lescanne_2020,Xu2022PhysRevR,Chamberland_2022}. Because of these circumstances, analyzing quantum channels with biased noise forms an important area in fault-tolerant quantum computation. Error-correcting codes optimized for biased-noise channels can improve the error threshold compared to conventional ones~\cite{Aliferis_2008,R_thlisberger_2012,Brooks_2013,Stephens_2013,Tuckett_2018,Li_2019,Tuckett_2020,Bonilla_Ataides_2021}, stimulating not only theoretical but also practical interest.

This paper examines the viability of fault-tolerant quantum computation in biased noise channels, particularly focusing on Noisy Intermediate-Scale Quantum (NISQ) or early Fault-Tolerant Quantum Computational (FTQC) devices, which are equipped with many qubits but have limited depth.
Specifically, we focus on the repetition code, the simplest error-correcting code that would be useful in a highly biased environment and is feasible with modern technology~\cite{Reed_2012,Kelly_2015,Rist__2015,Wootton_2018,Chen2021,Livingston_2022}.
To clarify the setup, we will consider a noise model characterized exclusively by uncorrelated bit-flip errors; we will ignore coherent, readout, and other error types. The following argument holds true for a noise model in which only phase-flip errors occur.
The question here is whether a universal quantum computation is possible in such a setup.
This is highly nontrivial: despite the error channel being restricted to bit-flip errors, it remains possible for phase-flip errors to arise during the course of the computation, which cannot be detected nor corrected by the bit-flip repetition code. 
For example, a naively implemented Hadamard gate converts the bit-flip error to the phase-flip error.
Consequently, achieving fault-tolerant quantum computation demands the implementation of logical gates that preserve the bias of the error channel. Such gates, known as bias-preserving gates~\cite{Guillaud_2019,Puri_2020,Xu2022PhysRevR}, must be designed to avoid converting bit-flip errors into other types of errors.

To address this issue, we propose a set of novel bias-preserving gates: $S$, $H$, CZ, and $R_z$ gates. Our central concept is that phase-flip errors occurring in the middle of the circuits can be intentionally correlated with bit-flip errors. Consequently, this correlation enables the detection of the phase-flip errors emerging within these Clifford gates through usual syndrome measurements. While errors in the $R_z$ gate can no longer be Pauli errors, the bias-preserving $R_z$ gate can be implemented based on the same idea. 
While our implementation approach entails the drawback of requiring postselection, it offers the advantage of bypassing the need for the Toffoli gate in Hadamard gate implementation~\cite{Guillaud_2019}. 
Another previous study proposed a bias-preserving implementation for a class of operators called $X$-type unitaries~\cite{fellousasiani2023scalable}. In contrast, a notable feature of this study is incorporating analog rotation gates that allow arbitrary unitary operations.

This paper is organized as follows. 
In Sec.~\ref{Sec:gates}, we define our noise model and show the details of implementing fault-tolerant and bias-preserving gates.
In Sec.~\ref{Sec:demo}, we show several numerical demonstrations of our scheme.
In Sec.~\ref{sec:phaseflip}, we discuss the effects of phase-flip error.
Section~\ref{sec:discussion} is devoted to the discussion and concluding remarks.

\section{Fault-tolerant and bias-preserving logical gates}\label{Sec:gates}
Throughout this paper, $X$, $Y$ and $Z$ denote Pauli matrices. The Pauli matrix $P$ that acts on $i$-th qubit is denoted by $P_i \equiv I \otimes \cdots \otimes P \otimes \cdots \otimes I$, where $I$ is the identity matrix.
The eigenstates of $X$ and $Y$ gates are denoted by $\ket{\pm}$ and $\ket{\pm i}$, respectively, i.e. $X\ket{\pm}=\pm\ket{\pm}$ and $Y\ket{\pm i}=\pm\ket{\pm i}$.

First, we define our noise model and clarify the meaning of fault-tolerance and bias-preserving gates in Sec.~\ref{Sec:noise model}.
In the next section~\ref{Sec:error-correction}, we remind well-known fault-tolerant error-correcting circuit and a slightly modified version.
In the remaining sections~\ref{Sec:Clifford gates} and~\ref{Sec:rotation gates}, we propose fault-tolerant and bias-preserving logic gates.

\subsection{The noise model and fault-tolerant operations}\label{Sec:noise model}
We discuss quantum circuits in a noisy environment where only bit-flip error occurs.
We assume that elemental gate sets consist of single and two-qubit gates.
Let $\rho_0$ and $\rho$ be the ideal initial and final density matrices of the system, and $U$ be a unitary gate. 
The state after a noisy single-qubit unitary operation is given by $\calE_p(U\rho_0U^\dagger) = \calE_p(\rho)$, where $\calE_p$ denotes the quantum operation that describes the bit-flip error after a single qubit with probability $p$:
\begin{align}
    \calE_p(\rho) = \sum_{j=0,1} E_j(p) \rho E_j^\dagger(p), \quad 
    E_0 = \sqrt{1-p}I, \quad 
    E_1 = \sqrt{p}X.
    \label{eq: bitflip noise channel}
\end{align}
The noisy two-qubit gate is given in a similar manner, but the noise channel is replaced by
\begin{align}
    (\calE_{p} \otimes \calE_{p})(\rho)
    &=
    \sum_{j=0}^3 E_j(p) \rho E_j^\dagger(p),  \\
    E_0(p) &= (1-p)I,  \\
    E_1(p) &= \sqrt{(1-p)p}X_1,  \\
    E_2(p) &= \sqrt{(1-p)p}X_2,  \\
    E_3(p) &= pX_1X_2.
\end{align}
Unless otherwise noted, we do not consider other types of errors: phase-flip error, coherent error, readout error and so on.
We also assume that we can always prepare clean $\ket{0}$
\footnote{Actual quantum devices may not satisfy this assumption. Even in that case, though, one can prepare the state in a fault-tolerant manner. A method is discussed in Appendix~\ref{app:state preparation}}.

Encoding qubits in the bit-flip code is the simplest way to protect qubits from the bit-flip error.
In the bit-flip code with distance $d=3$, we make each qubit redundant with three qubits as
\begin{align}
    \ket{\psi} \equiv a\ket{0} + b\ket{1}
    \longrightarrow
    \ket{\psi}_L \equiv
    a\ket{000} + b\ket{111}, \quad (|a|^2 + |b|^2 = 1, \, a,b \in \mathbb{C}).
    \label{eq: bit-flip code}
\end{align}
Note that the encoded state, or the logical qubit, is denoted as $\ket{\cdot}_L$ from now on. Each qubit that constitutes a logical qubit is called a physical qubit.
This code can correct a single-qubit bit-flip error but cannot correct further errors, which we call logic errors.

Let $N$ be the number of logical qubits, and $X_{a,j}$ be an $X$-gate acts on $j$-th physical qubit of the $a$-th logical qubit.
An operation is called a fault-tolerant and bias-preserving operation if, after performing the operation, the density matrix is given by
\begin{align}
    (1 - c_0p) \rho
    +
    \sum_{a=1}^N
    \sum_{j=1}^3
    c_{aj} p X_{a,j} \rho X_{a,j}^\dagger 
    +
    O(p^2),
    \label{eq: def of fault-tolerant operation}
\end{align}
where $\rho$ is the density matrix for the noiseless case, and $c_0$ and $c_{aj}$ are $O(1)$ constants.
The definition of the fault-tolerant operation requires logic errors do not occur with probability $O(p)$. In other words, terms like $pX_{1,1}X_{1,2}\rho X_{1,2}X_{1,1}$ are not allowed.
At the same time, Eq.~\eqref{eq: def of fault-tolerant operation} also requires that the phase-flip error does not appear with probability $O(p)$. 
As we will see below, the phase-flip error inevitably appears during several logical gates, even for our noise model, if we implement these logical gates naively.
Constructing bias-preserving logical gates, which only cause a bit-flip error, is a central issue of the following sections.

\subsection{Error correction}\label{Sec:error-correction}
\begin{figure}[h]
\begin{align*}
    \Qcircuit @C=2.em @R=2.em {
            &\ctrl{3}&\qw     &\qw     &\qw     &\qw   \gategroup{1}{2}{5}{6}{.7em}{--}&\qw&\gate{X}    &\qw\\
            &\qw     &\ctrl{2}&\ctrl{3}&\qw     &\qw                                   &\qw&\gate{X}\cwx&\qw\\
            &\qw     &\qw     &\qw     &\ctrl{2}&\qw                                   &\qw&\gate{X}\cwx&\qw\\
        \ini&\targ   &\targ   &\qw     &\qw     &\meter                                &\cw&\cw\cwx     &\\
        \ini&\qw     &\qw     &\targ   &\targ   &\meter                                &\cw&\cw\cwx     &\\
            & & &\mbox{syndrome measurement} & &\mbox{$\times n$} & & &
    }
\end{align*}
\caption{A quantum circuit for the fault-tolerant error correction. The part of the circuit enclosed by the dotted line forms the syndrome measurement. It should be repeated twice or more. The $X$-gates connected by double wires means the operating an $X$-gate to a physical qubit with a suspicious bit-flip error identified by the results of the syndrome measurement.}
\label{fig: error correction}
\end{figure}
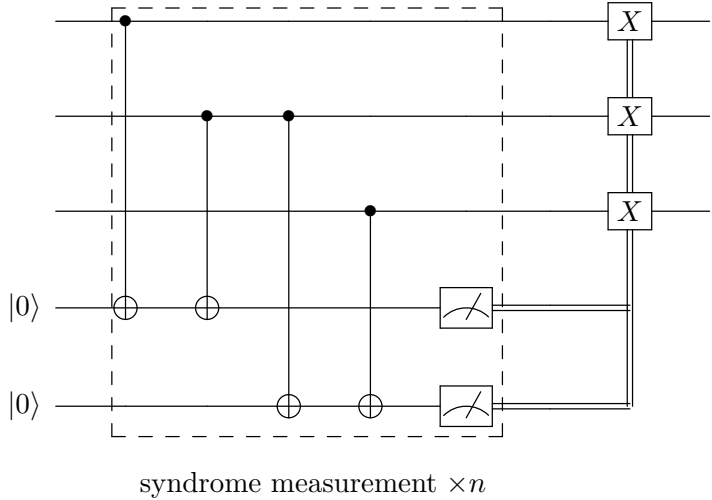

A bit-flip error in the code Eq.~\eqref{eq: bit-flip code} can be detected by the measurement of the stabilizers $Z_1Z_2$ and $Z_2Z_3$, and it can be corrected by the subsequent feedback operation of $X_i$. 
For completeness, we show the relation between the error syndrome and the feedback operation required in 
Table~\ref{tab: syndromes}. 
The eigenvalues of the stabilizers can be obtained by measuring ancilla qubits connected by CNOT gates with the logical qubit as shown in Fig.~\ref{fig: error correction}.

If the error syndrome is measured only once, bit-flip errors can accumulate in two physical qubits with probability $p$.
This event is caused by a bit-flip error after the second CNOT gate.
To evade this problem, it is well known that we need to take a majority vote on multiple syndrome measurements to obtain the correct error syndrome, known as Shor's fault-tolerant scheme~\cite{Shor548464}. Here, we revisit the three-round scheme~\cite{NielsenChuang}[Sec. 10.6]. If all three syndrome measurements differ from each other and the majority vote cannot be taken, no feedback is manipulated.
In this implementation, there are $24$ different failure patterns such that the probability of occurrence is $p$ within our noise model.
These errors can be recovered or give at most one bit-flip error in the final state.
The density matrix after the fault-tolerant error correction is
\begin{align}
    (1-8p)\rho
    +
    2p X_1 \rho X_1^\dagger
    +
    4p X_2 \rho X_2^\dagger
    +
    2p X_3 \rho X_3^\dagger
    +
    O(p^2).
\end{align}

This parametrization can be slightly improved by reducing the number of syndrome measurements to two.
Two syndrome measurements are sufficient to distinguish bit-flip errors occurring after the second CNOT gate, making the circuit non-fault-tolerant from other events.
Proof of this and a lookup table for the feedback operations are shown in Appendix~\ref{app:fault-tolerant error correction}. 
We find that the state after this shorthand error correction circuit is given by
\begin{align}
    (1-7p)\rho
    +
    3p X_1 \rho X_1^\dagger
    +
    2p X_2 \rho X_2^\dagger
    +
    2p X_3 \rho X_3^\dagger
    +
    O(p^2).
\end{align}
Below, we use this shorthand circuit to correct bit-flip errors.

\begin{table}[hbtp]
     \caption{In the first and second columns, we show the eigenvalue of $Z_1Z_2$ and $Z_2Z_3$. The feedback operation to correct the error is shown in the third column.}
     \centering
     \begin{tabular}{ccc}
     \hline
     $Z_1Z_2$ & $Z_2Z_3$ & feedback  \\
     \hline
      $1$ & $1$ & $I$ \\
     $-1$ & $1$ & $X_1$ \\   
     $-1$ & $-1$ & $X_2$ \\  
     $1$ & $-1$ & $X_3$ \\  
     \hline
     \end{tabular}
     \label{tab: syndromes}
\end{table}

\subsection{Clifford gates}\label{Sec:Clifford gates}
\begin{figure}
    \centering
    \begin{align*}
    \Qcircuit @C=2.em @R=2.em {
        &\gate{X}&\qw &&\gate{Y}&\qw &&\gate{Z}&\qw\\
        &\gate{X}&\qw &&\gate{X}&\qw &&\qw&\qw\\
        &\gate{X}&\qw &&\gate{X}&\qw &&\qw&\qw
    }
    \end{align*}
    \caption{Logical $X$, $Y$, and $Z$ gates.}
    \label{fig: pauli gates}
\end{figure}
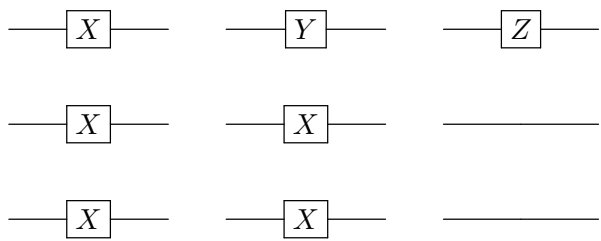

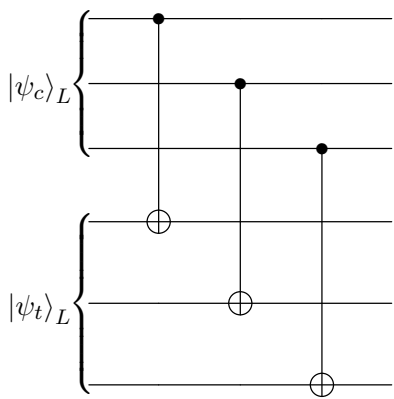
\begin{figure}
    \centering
    \begin{align*}
    \Qcircuit @C=2.em @R=2.em {
        &\ctrl{3}&\qw     &\qw     &\qw\\
        &\qw     &\ctrl{3}&\qw     &\qw\\
        &\qw     &\qw     &\ctrl{3}&\qw\\
        &\targ   &\qw     &\qw     &\qw\\
        &\qw     &\targ   &\qw     &\qw\\
        &\qw     &\qw     &\targ   &\qw
        \inputgroupv{1}{3}{.5em}{2.4em}{\ket{\psi_c}_L}
        \inputgroupv{4}{6}{.5em}{3em}{\ket{\psi_t}_L}
    }
    \end{align*}
    \caption{The fault-tolerant logical CNOT gate. The controlled and target logical qubits are denoted by $\ket{\psi_c}_L$ and $\ket{\psi_t}_L$, respectively.}
    \label{fig: cnot gate}
\end{figure}

Logical Pauli and CNOT gates are straightforwardly implemented as fault-tolerant gates.
They are shown in Figs.~\ref{fig: pauli gates} and~\ref{fig: cnot gate}.
On the other hand, the implementation of logical $S$ and $H$ gates are involved.
While the logical $S$ gate could naively be thought to be implemented as $S_L = S_1$, it is no longer bias-preserving. When it comes to the $H$ gate, it is not possible to implement it transversely.

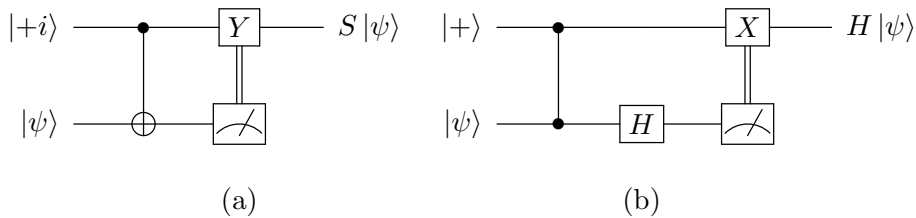
\begin{figure}
    \centering
    \begin{align*}
    \Qcircuit @C=2.em @R=2.em {
        \lstick{\ket{+i}}&\ctrl{1}&\gate{Y}  &\rstick{S\ket{\psi}}\qw &&&\iniX&\ctrl{1}    &\qw     &\gate{X}  &\rstick{
        H\ket{\psi}}\qw\\
        \lstick{\ket{\psi}} &\targ   &\meter\cwx&    &&&\lstick{\ket{\psi}}&\control \qw&\gate{H}&\meter\cwx& \\
        &&\mbox{(a)} &&&&&&\mbox{(b)}&&
    }
    \end{align*}
    \caption{(a) $S$ gate teleportation. (b) $H$ gate teleportation. In these figures, $Y$ and $X$ gates connected by double lines to the meter are activated only when the result $-1$ is obtained.}
    \label{fig: S and H gates}
\end{figure}

To evade these difficulties, we examine gate teleportation, which is an essential building block of fault-tolerant quantum computation~\cite{Gottesman_1999,GourlayPhysRevA.62.022308}.
One way to achieve $S$ and $H$ gates via teleportation is shown in Fig.~\ref{fig: S and H gates}. 
We note that the gate teleportation of $S^\dagger$ gate is obtained by replacing $\ket{+i}$ by $\ket{-i}$ in Fig.~\ref{fig: S and H gates} (a).
In both circuits, the Pauli gate connected to the meter by the double line is meant to operate when the result $-1$ is obtained.

Our goal is to prove each component in these circuits, state preparation, CZ gate, and measurement in the $X$ basis, can be implemented as a fault-tolerant and bias-preserving operation.

\subsubsection{Preparation of $\ket{+}_L$}
Within our noise model,
the eigenstate of logical $X$ can be prepared directly by the circuit in Fig.~\ref{fig: |+> preparation}.
Noting that CNOT gates do not amplify the bit-flip error since the error after the $H$ gate does not change the state ($X\ket{+}=\ket{+}$), the final state reads
\begin{align}
    (1-4p) \rho
    +
    p X_1 \rho X_1^\dagger 
    +
    2p X_2 \rho X_2^\dagger 
    +
    p X_3 \rho X_3^\dagger
    +
    O(p^2).
\end{align}
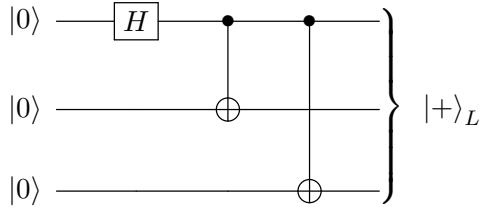
\begin{figure}
    \centering
    \begin{align*}
        \Qcircuit @C=2.em @R=2.em {
            \ini&\gate{H}&\ctrl{1}&\ctrl{2}&\qw\\
            \ini&\qw     &\targ   &\qw     &\rstick{\quad\ket{+}_L}\qw\\
            \ini&\qw     &\qw     &\targ   &\qw
            \gategroup{1}{5}{3}{5}{.8em}{\}}
        }
    \end{align*}
    \caption{A quantum circuit to prepare $\ket{+}_L$.}
    \label{fig: |+> preparation}
\end{figure}

\subsubsection{Preparation of $\ket{+i}_L$}
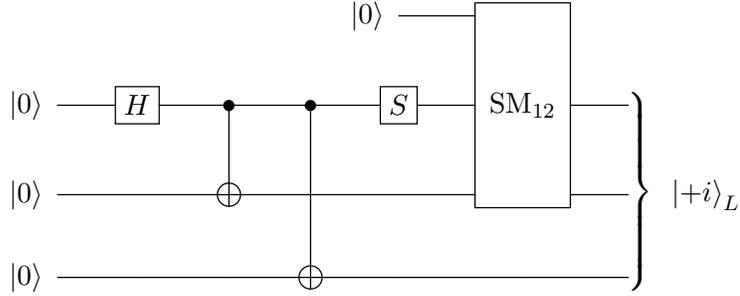
\begin{figure}
    \centering
    \begin{align*}
    \Qcircuit @C=2.em @R=2.em {
            &        &        &        &\ini    &\multigate{2}{\text{SM}_{12}}&\\
        \ini&\gate{H}&\ctrl{1}&\ctrl{2}&\gate{S}&\ghost{\text{SM}_{12}}       &\qw\\
        \ini&\qw     &\targ   &\qw     &\qw     &\ghost{\text{SM}_{12}}       &\rstick{\quad\ket{+i}_L}\qw\\
        \ini&\qw     &\qw     &\targ   &\qw     &\qw                     &\qw
        \gategroup{2}{7}{4}{7}{.8em}{\}}
    }
    \end{align*}
    \caption{A quantum circuit to prepare $\ket{+i}_L$. The SM$_{12}$ gate denotes the indirect measurement of $Z_1Z_2$.}
    \label{fig: |+i> preparation}
\end{figure}

The eigenstate of logical $Y$ is prepared by the circuit shown in Fig.~\ref{fig: |+i> preparation}.
This circuit consists of $\ket{+}_L$-preparation followed by the $S$ gate.
Note that $\ket{-i}_L$ can be obtained by replacing the $S$ gate by $S^\dagger$ gate in this circuit.
A bit-flip error on the first physical qubit before the $S$ gate is converted to a $Y$ error, and thus, it breaks the bias-preservation.
However, the possible $Y$ error can be detected by the measurement of the stabilizer $Z_1Z_2$ depicted by $\text{SM}_{12}$ in Fig.~\ref{fig: |+i> preparation}, and eliminated by postselection.
The output of this circuit after such postselection is given by
\begin{align}
    (1-3p)\rho
    +
    p X_1 \rho X_1^\dagger
    +
    p X_2 \rho X_2^\dagger
    +
    p X_3\rho X_3^\dagger
    +
    O(p^2).
\end{align}

\subsubsection{CZ gate}
The CZ gate is not bias-preserving due to the relation $\CZ_{ij}X_i = X_iZ_j\CZ_{ij}$.
On the other hand, the phase-flip error does not appear alone but together with a bit-flip error.
This feature enables us to detect the phase-flip error indirectly.
We show a fault-tolerant and bias-preserving logical CZ gate in Fig.~\ref{fig: CZ gate} that consists of a CZ gate between the first physical qubits followed by syndrome measurements. Note that if the ancilla can be recycled, only one ancilla is needed for the syndrome measurement, so only one $\ket{0}$ is depicted in the figure.
To see how this circuit works, assume that the incoming first logical qubit $\ket{\psi_1}_L$ has a bit-flip error on the first physical qubit.
Propagating the bit-flip error forward, a phase-flip error appears in the second logical qubit $\ket{\psi_2}_L$ while the bit-flip error remains. 
Thus, the syndrome measurement for the first logical qubit can detect the phase-flip error in the second logical qubit.
Eliminating such events by postselection, the final state reads
\begin{align}
    (1-10p) \rho
    +
    \sum_{a=1}^2
    \left(
        2X_{a,1} \rho X_{a,1}^\dagger 
        +
        2X_{a,2} \rho X_{a,2}^\dagger 
        +
        X_{a,3} \rho X_{a,3}^\dagger 
    \right)
    +
    O(p^2) .
\end{align}
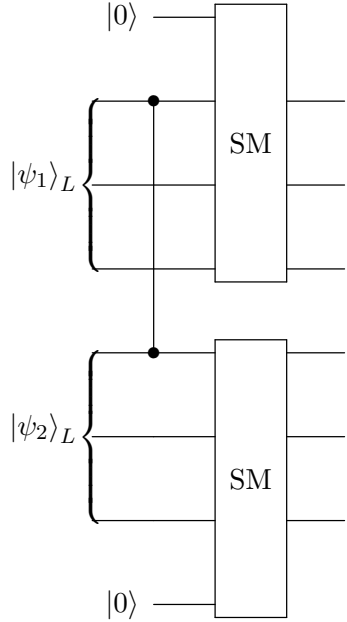
\begin{figure}
    \centering
    \begin{align*}
    \Qcircuit @C=2.em @R=2.em {
        &\ini       &\multigate{3}{\text{SM}}& \\
        &\ctrl{3}   &\ghost{\text{SM}}       &\qw\\
        &\qw        &\ghost{\text{SM}}       &\qw\\
        &\qw        &\ghost{\text{SM}}       &\qw\\
        &\control\qw&\multigate{3}{\text{SM}}&\qw\\
        &\qw        &\ghost{\text{SM}}       &\qw\\
        &\qw        &\ghost{\text{SM}}       &\qw
        \inputgroupv{2}{4}{.1em}{2.6em}{\ket{\psi_1}_L}
        \inputgroupv{5}{7}{.1em}{2.6em}{\ket{\psi_2}_L}\\
        &\ini       &\ghost{\text{SM}}       & 
    }
    \end{align*}
    \caption{A fault-tolerant and bias-preserving logical CZ gate. The first and second logical qubits are denoted by $\ket{\psi_1}_L$ and $\ket{\psi_2}_L$, respectively. The SM gate denotes the syndrome measurement.}
    \label{fig: CZ gate}
\end{figure}

\subsubsection{Measurement in the $X$ basis}
\begin{figure}
    \centering
    \begin{align*}
    \Qcircuit @C=2.em @R=2.em {
          &\qw&\qw  &\targ    &\qw     &\qw\gategroup{1}{3}{4}{5}{.7em}{--}\\
          &\qw&\qw  &\targ    &\qw     &\qw\\
         &\qw&\qw  &\targ    &\qw     &\qw\\
        &\ini&\gate{H}&\ctrl{-3}&\gate{H}&\meter&\\
        &&&&&\mbox{$\times 3$}
    }    
    \end{align*}
    \caption{A quantum circuit to measure a logical state in the $X$ basis.}
    \label{fig: X measurement}
\end{figure}
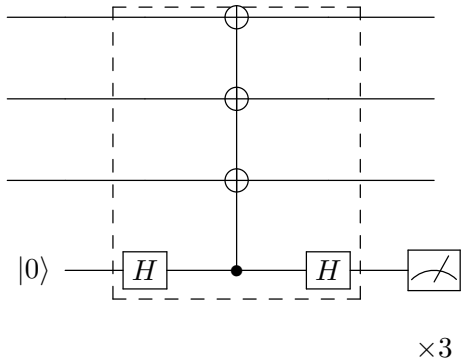

Measurement of the logical state in the $X$ basis can be performed by the circuit shown in Fig.~\ref{fig: X measurement}, where the set of gates enclosed by the dotted line should be repeated three times.
The probability of obtaining an incorrect measurement is $O(p^2)$ by taking a majority vote on the results of three measurements.

The probability that an uncorrectable error accumulates on the output state is also $O(p^2)$.
This can be seen from the fact that any error in the ancilla is not spread over more than two physical qubits by noting that the bit-flip error does not change $\ket{+}$.

\subsubsection{$S$ gate}
\begin{figure}
    \centering
    \begin{align*}
    \Qcircuit @C=2.em @R=2.em {
        \ini&\multigate{3}{\ket{+i}_L\text{-prep.}}&        &        &        &            &\\
        \ini&\ghost{\ket{+i}_L\text{-prep.}}       &\ctrl{3}&\qw     &\qw     &\gate{Y}    &\qw\\
        \ini&\ghost{\ket{+i}_L\text{-prep.}}       &\qw     &\ctrl{3}&\qw     &\gate{X}\cwx&\rstick{\quad S_L\ket{\psi}_L}\qw\\
        \ini&\ghost{\ket{+i}_L\text{-prep.}}       &\qw     &\qw     &\ctrl{3}&\gate{X}\cwx&\qw
        \inputgroupv{5}{7}{.1em}{3.4em}{\ket{\psi}_L}\gategroup{2}{7}{4}{7}{.8em}{\}}\\ 
            &\qw                                   &\targ   &\qw     &\qw     &\meter\cwx  &\\
            &\qw                                   &\qw     &\targ   &\qw     &\meter\cwx  &\\
            &\qw                                   &\qw     &\qw     &\targ   &\meter\cwx  &\\
    }
    \end{align*}
    \caption{A quantum circuit of logical $S$ gate teleportation.}
    \label{fig: logical S gate}
\end{figure}
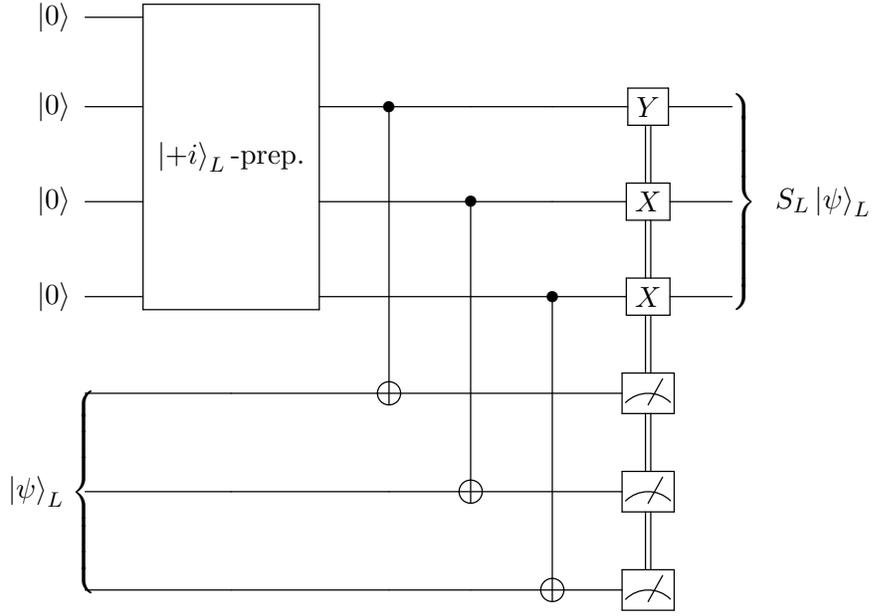

Encoding each component in Fig.~\ref{fig: S and H gates} (a), we obtain a quantum circuit of logical $S$ gate teleportation.
We show the entire circuit in Fig.~\ref{fig: logical S gate}.
Note that the preparation of $\ket{+i}_L$ needs appropriate postselection to eliminate phase-flip errors as we saw before. 
However, if we allow a repeat-until-success approach to prepare the state, the postselection is no longer mandatory.
The logical $S$ gate teleportation requires seven qubits in total: one logical data qubit, one logical resource qubit, and one ancilla if the ancilla can be recycled.
The parametrization of the output state depends on the result of the measurement of the logical data qubit.
If ancilla is $\ket{0}$ (in the sense of the majority vote), the output state reads
\begin{align}
    (1-6p) \rho
    +
    2p X_1 \rho X_1^\dagger
    +
    2p X_2 \rho X_2^\dagger
    +
    2p X_3 \rho X_3^\dagger
    +
    O(p^2).
\end{align}
If $\ket{1}$ is measured and the logical $Y$ gate is operated, 
we obtain
\begin{align}
    (1-9p) \rho
    +
    3p X_1 \rho X_1^\dagger
    +
    3p X_2 \rho X_2^\dagger
    +
    3p X_3 \rho X_3^\dagger
    +
    O(p^2).
\end{align}

\subsubsection{$H$ gate}
\begin{figure}
    \centering
    \begin{align*}
    \Qcircuit @C=2.em @R=2.em {
                &                                     &\ini       &\multigate{3}{\text{SM}}&    &                                 & \\
            \ini&\multigate{2}{\text{prep.-}\ket{+}_L}&\ctrl{3}   &\ghost{\text{SM}}       &\qw &\gate{X}                         &\qw\\
            \ini&\ghost{\text{prep.-}\ket{+}_L}       &\qw        &\ghost{\text{SM}}       &\qw &\gate{X}\cwx                     &\rstick{\quad H_L\ket{\psi}_L}\qw\\
            \ini&\ghost{\text{prep.-}\ket{+}_L}       &\qw        &\ghost{\text{SM}}       &\qw &\gate{X}\cwx                     &\qw
            \inputgroupv{5}{7}{.1em}{3.4em}{\ket{\psi}_L}\gategroup{2}{7}{4}{7}{.8em}{\}}\\
                &\qw                                  &\control\qw&\multigate{3}{\text{SM}}&\qw &\multigate{3}{X\text{-meas.}}\cwx& \\
                &\qw                                  &\qw        &\ghost{\text{SM}}       &\qw &\ghost{X\text{-meas.}}           & \\
                &\qw                                  &\qw        &\ghost{\text{SM}}       &\qw &\ghost{X\text{-meas.}}           & \\
                &                                     &\ini       &\ghost{\text{SM}}       &\ini&\ghost{X\text{-meas.}}           & 
    }
    \end{align*}
    \caption{A quantum circuit of logical $H$ gate teleportation.}
    \label{fig: logical H gate}
\end{figure}
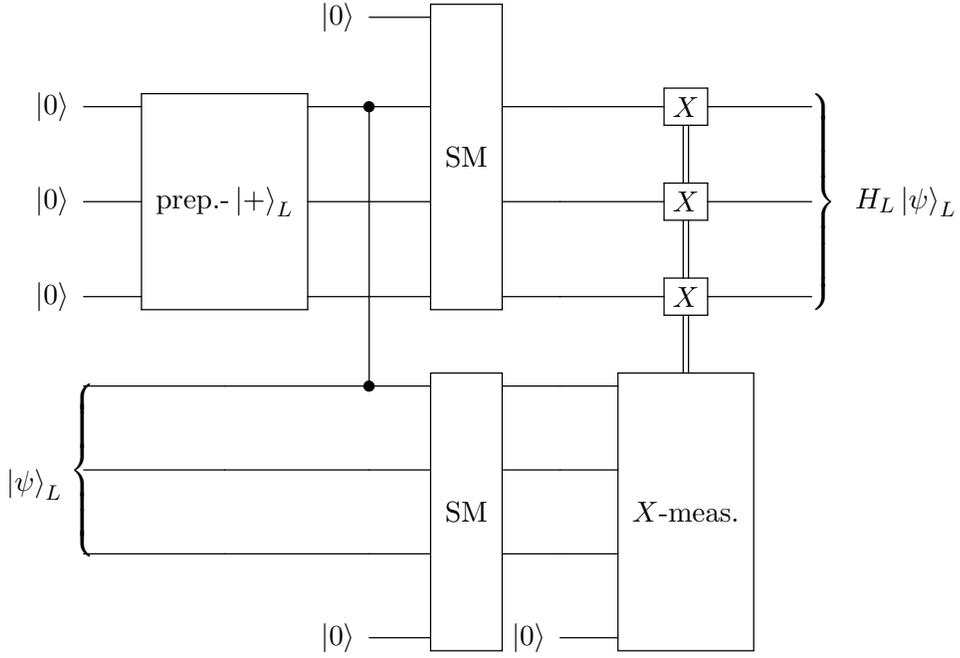

Encoding each component in Fig.~\ref{fig: S and H gates} (b), we obtain a quantum circuit of logical $H$ gate teleportation.
We show the entire circuit in Fig.~\ref{fig: logical H gate}.
This circuit needs postselection to eliminate phase-flip errors that can appear after the CZ gate.
The logical $H$ gate teleportation requires at least seven qubits, assuming that ancilla can be recycled.
The parameterization of the output state depends on the result of the measurement of the logical data qubit.
If $\ket{0}_L$ is observed, the output state reads
\begin{align}
    (1-9p) \rho
    +
    3p X_1 \rho X_1^\dagger
    +
    4p X_2 \rho X_2^\dagger
    +
    2p X_3 \rho X_3^\dagger
    +
    O(p^2),
\end{align}
otherwise
\begin{align}
    (1-12p) \rho
    +
    4p X_1 \rho X_1^\dagger
    +
    5p X_2 \rho X_2^\dagger
    +
    3p X_3 \rho X_3^\dagger 
    +
    O(p^2).
\end{align}

\subsection{Rotation gates}\label{Sec:rotation gates}
We discuss the implementation of rotation gates, namely the $R_z$ gate.
Once the $R_z$ gate is made, the rotation gate in other bases can be constructed using fault-tolerant Clifford gates.
We start by pointing out that the $R_z$ gate can be operated indirectly, as shown in Fig.~\ref{fig: indirect Rz gate}.
The conditional $Z$ gate acts on the target state only when the result $-1$ is obtained.

\begin{figure}
    \centering
    \begin{align*}
    \Qcircuit @C=2.em @R=2.em {
            &\ctrl{3}&\qw     &\qw     &\qw               &\qw                     &\qw &\qw                              &\qw\\
            &\qw     &\ctrl{3}&\qw     &\qw               &\qw                     &\qw &\qw                              &\rstick{\quad R_z(\theta)_L\ket{\psi}_L}\qw\\
            &\qw     &\qw     &\ctrl{3}&\qw               &\qw                     &\qw &\gate{Z}                         &\qw
            \inputgroupv{1}{3}{.1em}{2.4em}{\ket{\psi}_L}\gategroup{1}{9}{3}{9}{.8em}{\}}\\
        \ini&\targ   &\qw     &\qw     &\gate{R_z(\theta)}&\multigate{3}{\text{SM}}&\qw &\multigate{3}{X\text{-meas.}}\cwx& \\
        \ini&\qw     &\targ   &\qw     &\qw               &\ghost{\text{SM}}       &\qw &\ghost{X\text{-meas.}}           & \\
        \ini&\qw     &\qw     &\targ   &\qw               &\ghost{\text{SM}}       &\qw &\ghost{X\text{-meas.}}           & \\
            &        &        &        &\ini              &\ghost{\text{SM}}       &\ini&\ghost{X\text{-meas.}}           & 
    }
    \end{align*}
    \caption{A quantum circuit of the fault-tolerant logical $R_z$ gate.}
    \label{fig: logical Rz gate}
\end{figure}
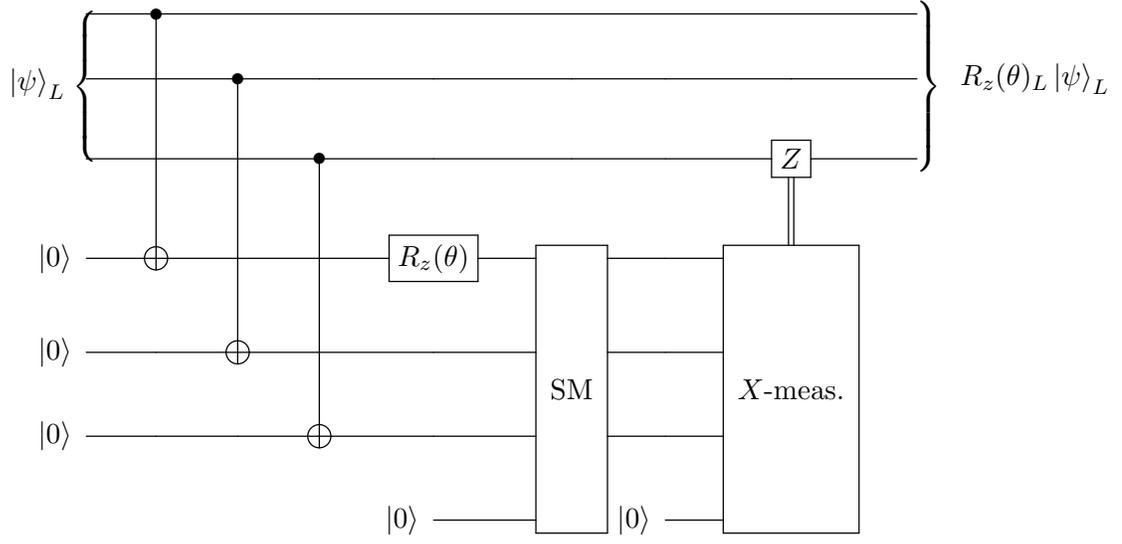

To obtain the fault-tolerant version of this circuit, we employ the same strategy as the fault-tolerant $S$ and $H$ gates construction, detecting unacceptable events by the syndrome measurement.
In this case, the event we should detect is that a bit-flip error arises before the $R_z$ gate. Such an error changes the sign of the rotation angle since $R_z(\theta)X = XR_z(-\theta)$.
If this happens, the encoded ancilla will be $\ket{100}$, which the syndrome measurement can detect.
Thus, we reach the circuit shown in Fig.~\ref{fig: logical Rz gate}.
This circuit needs at least seven qubits if the ancilla can be recycled.
The output state is
\begin{align}
    (1-3p) \rho
    +
    p X_1 \rho X_1^\dagger
    +
    p X_2 \rho X_2^\dagger
    +
    p X_3 \rho X_3^\dagger
    +
    O(p^2),
\end{align}
if 0 is measured in the ancilla side, otherwise
\begin{align}
    (1-4p) \rho
    +
    p X_1 \rho X_1^\dagger
    +
    p X_2 \rho X_2^\dagger
    +
    2p X_3 \rho X_3^\dagger 
    +
    O(p^2).
\end{align}
\begin{figure}
    \centering
    \begin{align*}
    \Qcircuit @C=2.em @R=2.em {
        \lstick{\ket{\psi}}&\ctrl{1}&\qw               &\qw     &\gate{Z}  &\rstick{R_z(\theta)\ket{\psi}}\qw\\
        \ini&\targ   &\gate{R_z(\theta)}&\gate{H}&\meter\cwx&
    }
    \end{align*}
    \caption{A quantum circuit to operate $R_z$ gate.}
    \label{fig: indirect Rz gate}
\end{figure}
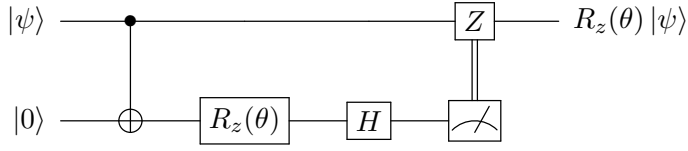

\section{Numerical demonstrations}\label{Sec:demo}
We apply our scheme to several problems to demonstrate the effectiveness of fault-tolerant computation in a noisy environment.

\subsection{Benchmark}
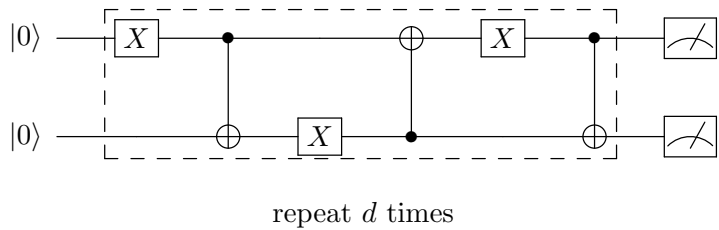
\begin{figure}
    \centering
    \begin{align*}
        \Qcircuit @C=2.em @R=2.em {
            \ini & \gate{X} & \ctrl{1} & \qw      & \targ     & \gate{X} & \ctrl{1} & \meter \\ 
            \ini & \qw      & \targ    & \gate{X} & \ctrl{-1} & \qw      & \targ    & \meter
            \gategroup{1}{2}{2}{7}{.7em}{--} \\
            &&&\mbox{\quad \quad \quad repeat $d$ times}&&&&
        }
    \end{align*}
    \caption{A benchmark circuit.}
    \label{fig: benchmark circuit}
\end{figure}

Let us consider a benchmark circuit as shown in Fig.~\ref{fig: benchmark circuit}, which consists of $X$ and CNOT gates.
We repeat the part of the circuit enclosed by the dotted line $d$ times.
One can easily verify that the final state will be $\ket{00}$ regardless of  $d$ in a noiseless environment.
To quantify the effect of bit-flip noise, we define the average squared fidelity $F$ by
\begin{align}
    F = \frac{1}{N} \sum_{i=1}^N |\bra{00}\ket{\psi_i}|^2,
\end{align}
where $N$ denotes the number of shots and $\ket{\psi_i}$ denotes $i$-th sampled state vector.
For comparison, we also consider the following circuits.
\begin{enumerate}
    \item An encoded circuit: Logically same as that shown in Fig.~\ref{fig: benchmark circuit}, but each qubit is encoded by the bit-flip code.
    \item An encoded circuit with error correction: Insert $d$ error-correction circuits to the end of each layer of the above-mentioned circuit.
\end{enumerate}

\begin{figure}
    \centering
    \includegraphics{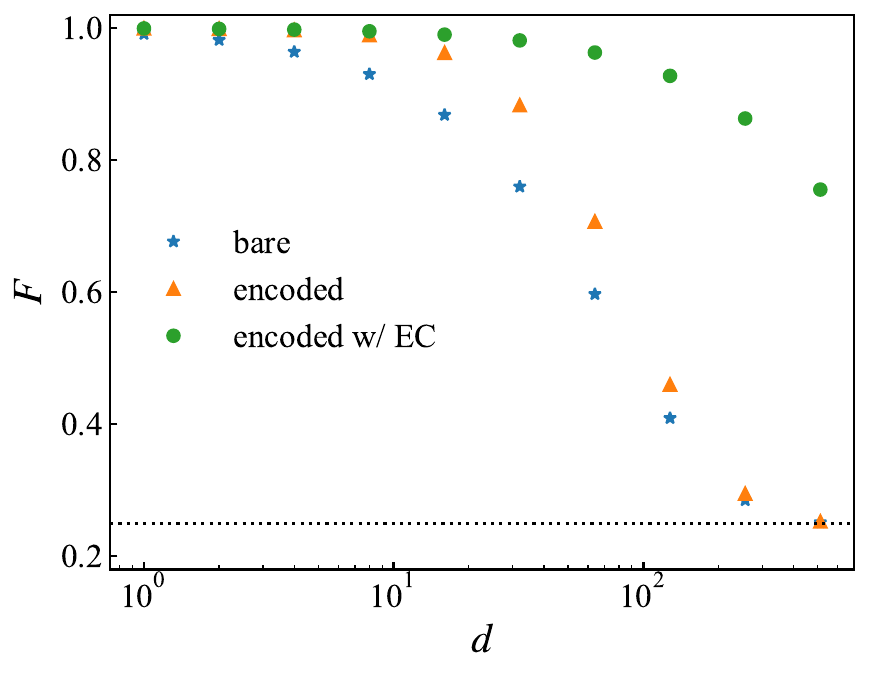}
    \caption{Average squared fidelity of the benchmark circuits. The dotted line indicates $F=1/4$, which means that the bit-flip noise totally randomizes the circuit. The star, triangle, and circle symbols correspond to the average fidelity of bare, encoded circuits, and encoded circuits with EC.}
    \label{fig: benchmark result}
\end{figure}

We perform numerical simulations of these circuits using Qiskit.
We set the gate error to $p=10^{-3}$.
We consider $d=2^0, 2^1, \dots, 2^9$ for each circuit and take $10^5$ shots for each simulation.
In Fig.~\ref{fig: benchmark result}, we show the average squared fidelity as a function of the depth for the bare (non-encoded), encoded circuits, and the circuit equipped with the error correction (EC).
While the encoded circuit without EC improves the scaling of the fidelity from $1-O(p)$ to $1-O(p^2)$, it shows poor performance as the depth increases. In particular, at $d=512$, it is almost entirely randomized, i.e., $F=1/4$ indicated by the dotted line.
In contrast, the encoded circuit with EC keeps fidelity high even at $d=512$.
This result clearly shows that the error-correction is mandatory in deep circuits.


\subsection{Dynamics of quantum system}
\begin{figure}
    \centering
    \begin{align*}
    \Qcircuit @C=2.em @R=2.em {
        &\ctrl{1}&\qw                &\ctrl{1}&\gate{R_x(2h\delta)}&\targ    &\gate{R_z(2\delta)}&\targ    &\qw & \qw\\
        &\targ   &\gate{R_z(2\delta)}&\targ   &\qw                 &\ctrl{-1}&\qw                &\ctrl{-1}&\gate{R_x(2h\delta)}&\qw
    }
    \end{align*}
    \caption{Repeating this circuit $M$ times gives the Trotterized time evolution operator Eq.~\eqref{eq: Trotterized time evolution operator} for $N=2$.} 
    \label{fig: time evolution circuit}
\end{figure}
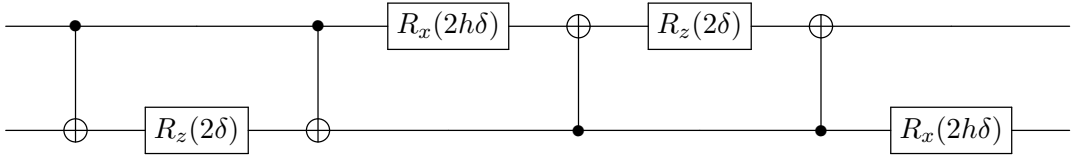

We consider the time evolution of a spin-chain model as an example of a deep nontrivial circuit.
Let us consider the transverse field Ising model 
\begin{align}
    \hat{H} = \sum_{i=1}^N (Z_iZ_{i+1} + hX_i)
\end{align}
with the periodic boundary condition $Z_{N+1}=Z_1$.
The time evolution of this model is approximated by the Suzuki-Trotter decomposition as
\begin{align}
    U = e^{-i\hat{H}t}
    \simeq
    \left(
        \prod_{i=1}^N e^{-iZ_iZ_{i+1}\frac{t}{M}}
        \prod_{i=1}^N e^{-iX_i\frac{t}{M}}
    \right)^M.
    \label{eq: Trotterized time evolution operator}
\end{align}
Below, we set $N=2$ for simplicity.
The time evolution operator is implemented as shown in Fig.~\ref{fig: time evolution circuit}, where $\delta = t/M$.
Note that ten physical qubits are needed to make this circuit fault-tolerant.

As for the previous subsection, we measure the performance of this circuit in a noisy environment.
For comparison, we also consider the encoded circuits with or without EC.
The fault-tolerant $R_x$ gate is implemented by decomposing the $R_x$ gate as $R_x = H R_z H$.
The EC is inserted every time after each fault-tolerant Hadamard gate.
We do not use such decomposition for the bare circuit.
We compute the expectation value of the total magnetization $M(t) = \sum_{i=1}^N \mel{\psi(t)}{Z_i}{\psi(t)}$ where $\ket{\psi(t)} = U\ket{00}$ taking $10^4$ shots at each time under the transverse magnetic field $h=1$.
We set the Trotter time slice to $\delta = 0.1$ and the gate error to $p=10^{-3}$.

\begin{figure*}[tb]
  \begin{subfigure}[t]{.45\textwidth}
    \centering
    \includegraphics[width=\linewidth]{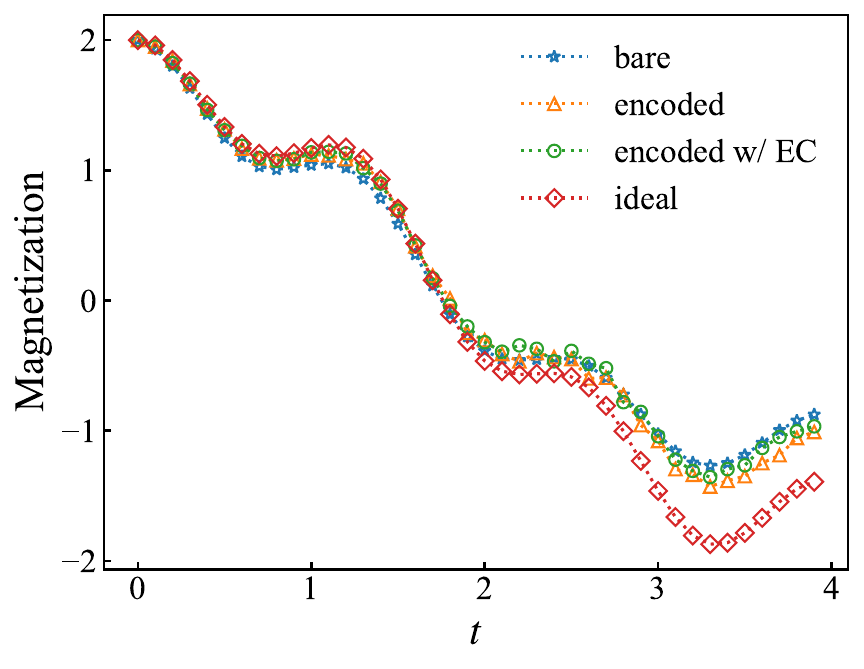}
    \caption{Magnetization}
    \label{fig: magnetization}
  \end{subfigure}
  \hfil
  \begin{subfigure}[t]{.45\textwidth}
    \centering
    \includegraphics[width=\linewidth]{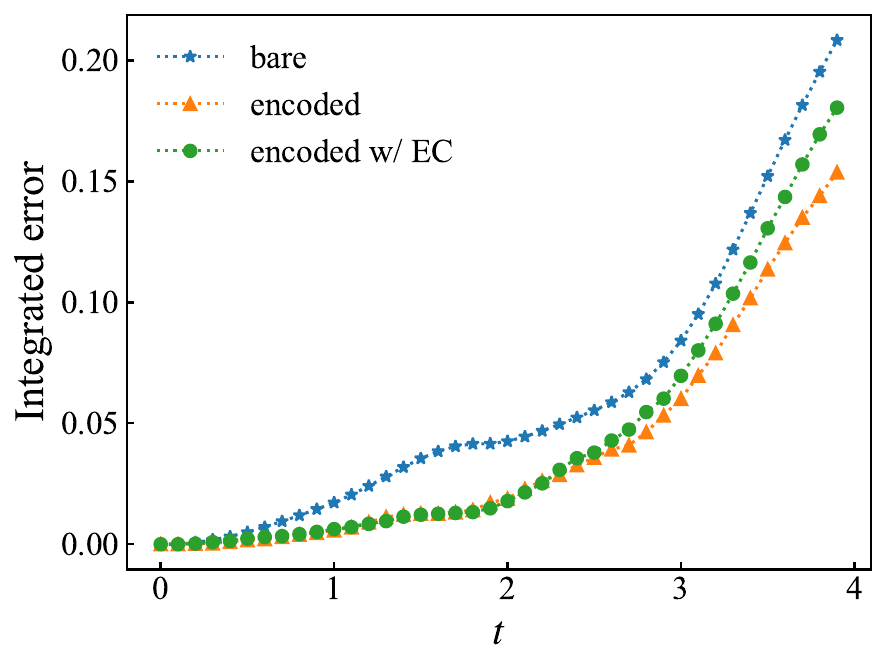}
    \caption{Integrated error}
    \label{fig: integrated error}
  \end{subfigure}
\medskip
  \begin{subfigure}[t]{.45\textwidth}
    \centering
    \includegraphics[width=\linewidth]{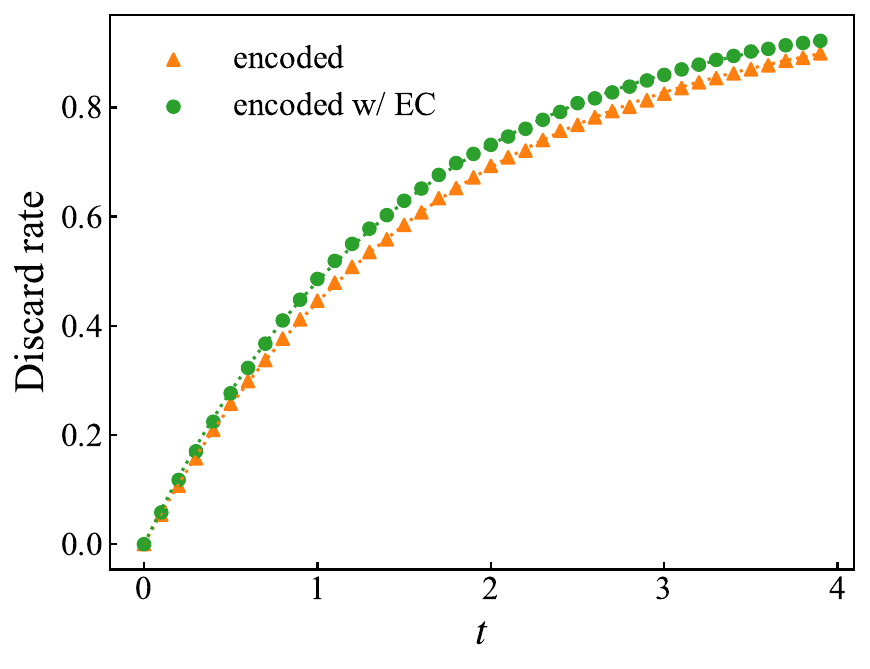}
    \caption{Discard rate}
    \label{fig: discard rate}
  \end{subfigure}
  \caption{Time evolution of the transverse Ising model. The diamond symbols indicate the noise-free results. Other symbols are the same as Fig.~\ref{fig: benchmark result}.}
  \label{fig: ising result}
\end{figure*}

In Fig.~\ref{fig: magnetization}, we show the total magnetization obtained from each quantum circuit at each Trotter step.
The results of all circuits appear to deviate from noiseless one.
To quantify the accumulation of errors, we define the integrated error by
\begin{align}
    E(N_\text{trot})
    =
    \frac{1}{N_\text{trot}}
    \sum_{i=1}^{N_\text{trot}}
    |M(i\delta) - M_{\text{noisy}}(i\delta)|,
\end{align}
where $M_{\text{noisy}}$ is the magnetization obtained by noisy quantum simulations and $N_\text{trot}$ is the number of Trotter steps.
We show the evolution of the integrated error in Fig.~\ref{fig: integrated error}.
On a quantitative level, the encoded circuits have a clear advantage up to $t \sim 2$.
In contrast to the case of the benchmark circuit consisting only of $X$ and CNOT gates, the EC circuit does not reduce the integrated error compared with the encoded circuit without EC.
In the encoded circuit, $Z$ errors that occur with probability $O(p)$ are removed by post-selection, but those that occur with probability lower than $O(p^2)$ remain unremoved.
Figure~\ref{fig: integrated error} suggests that the contribution of such uncorrectable errors is dominant in this simulation.

Finally, we discuss the yield of postselection. 
In Fig.~\ref{fig: discard rate}, we show the time dependence of the discard rate of postselection needed for fault-tolerant $H$ and $R_z$ gates.
In both cases, the discard rate exponentially approaches one.
According to the fitting by the function $1 - e^{-at}$ with a single parameter $a$, it gives $a \simeq 0.59$ for the encoded circuit, and $a \simeq 0.66$ for the encoded circuit with EC.
Post-selection exponentially worsens the efficiency of sampling, making it unsuitable for running deep circuits. However, if the time is limited to $t \lesssim 2$, which benefits from encoding, the discard rate is less than about 70\%.

\subsection{Variational quantum eigensolver}\label{sec: VQE}
We discuss the effectiveness of fault-tolerant computation in a shallow circuit.
To see this, we consider variational quantum eigensolver (VQE) using a hardware-efficient ansatz and apply it to quantum chemistry problems.
In general, the Hamiltonian of some molecule has a form of $\hat{H}=h_0+\sum_{pq} h_{pq}\hat{a}^\dagger_p\hat{a}_q + \sum_{pqrs}h_{pqrs}\hat{a}^\dagger_p\hat{a}^\dagger_q\hat{a}_r\hat{a}_s$, where $\hat{a}^{(\dagger)}_p$ is the annihilation (creation) operator of an electron in a molecular orbital $p$, $h_0$ is the nuclear repulsion energy, $h_{pq}$ is the one-electron integral and $h_{pqrs}$ is the two-electron integral. The electron integrals are obtained by the Hartree-Fock method, for instance.

In our demonstration, we aim to compute the ground state energy of caffeine (\ce{C8H10N4O2}) with restricted active space and the eigenspace of spin so that the Hamiltonian of a two-qubit system is obtained after the Jordan-Wigner transformation.
The concrete form of the qubit Hamiltonian and its derivation is presented in Appendix~\ref{app:caffeine hamiltonian}.
Since all coefficient of the Hamiltonian is real, it is sufficient to consider the following ansatz for VQE:
\begin{align*}
    \Qcircuit @C=2.em @R=2.em {
        \ini&\gate{R_y(\theta_1)}&\ctrl{1}   &\gate{R_y(\theta_3)}&\qw \\
        \ini&\gate{R_y(\theta_2)}&\control\qw&\gate{R_y(\theta_4)}&\qw 
    }
\end{align*}
where $\{\theta_i\}$ are the real parameters.
The fault-tolerant $R_y$ gate is implemented by decomposing the $R_y$ gate to $R_y = SH R_z HS^\dagger$.

We optimize these parameters classically and compute the ground state energy using the quantum circuit with those parameters.
The qubit Hamiltonian of the caffeine is divided into four groups of Pauli strings, and $10^7$ shots are taken to measure the expectation value of each. 
The gate error is $p=10^{-3}$ as before.
The result is summarized in Table~\ref{table: caffeine bit-flip error}, indicating that the encoded circuits reproduce the exact value well. On the other hand, the results are not so different from those of the bare circuit, so the benefit of encoding would be realized when better accuracy than chemical accuracy ($1\text{kcal/mol}\simeq 1.6\text{mHa}$) is required.
\begin{table}[hbtp]
    \caption{The ground state energy of caffeine.}
    \label{table: caffeine bit-flip error}
    \centering
    \begin{tabular}{lc}
    \hline
    Circuit type & Energy [Hartree]   \\
    \hline
    exact & $-667.7400$   \\ 
    bare & $-667.7394$  \\
    encoded & $-667.7397$ \\
    encoded with EC & $-667.7396$ \\
    \hline
    \end{tabular}
\end{table}

\section{Effect of phase-flip error}\label{sec:phaseflip}
From a practical standpoint, how circuits encoded with bit-flip codes behave in a general error channel is a question worth investigating.
Here, we perform a numerical simulation of VQE with the following error channel
\begin{align}
    \calE_{p,\epsilon}(\rho) 
    =
    (1-p)\rho 
    + 
    p(1-\epsilon)X^\dagger \rho X
    + 
    \frac{p\epsilon}{2}(
        Y^\dagger \rho Y
        +
        Z^\dagger \rho Z
    ).
\end{align}
We note that $\epsilon=0$ corresponds to the bit-flip noise channel Eq.~\eqref{eq: bitflip noise channel} and $\epsilon=2/3$ corresponds to the depolarizing channel.
Other setup details are the same as Sec.~\ref{sec: VQE} except the number of shots to calculate an expectation value of each Pauli group is $10^6$.

\begin{figure}
    \centering
    \includegraphics{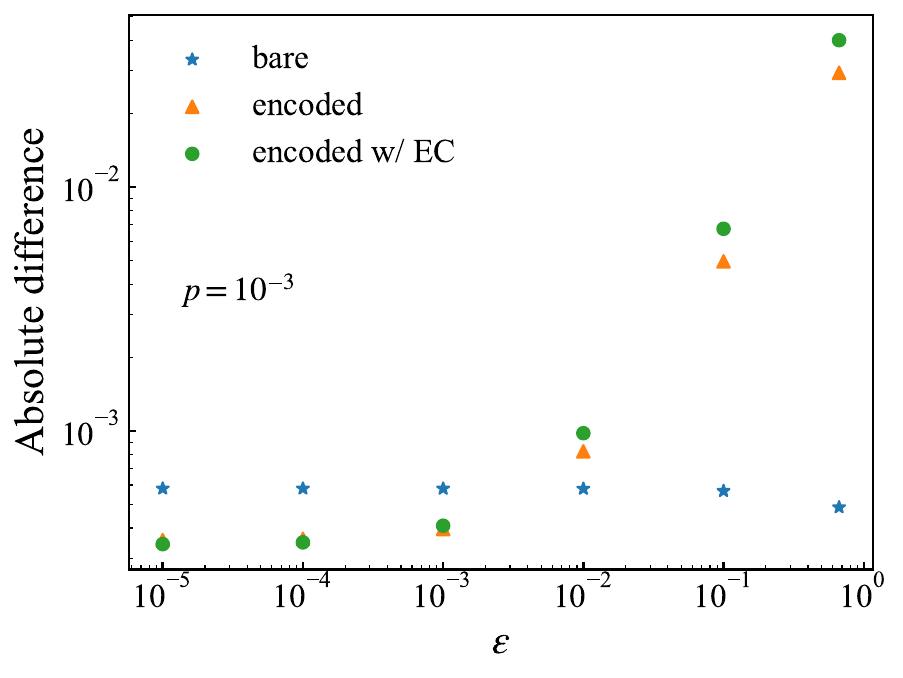}
    \caption{The absolute energy difference between exact and numerical results of caffeine. The symbols are the same as Fig.~\ref{fig: benchmark result}. The probability of the bit-flip error is $p=10^{-3}$.}
    \label{fig: Z error VQE}
\end{figure}

In Fig.~\ref{fig: Z error VQE}, we show the absolute difference between the exact ground state energy and numerical ones as a function of the magnitude of the $Z$ error.
The encoded circuits are superior to the bare one when $\epsilon = 10^{-3} = p$ or smaller. 
This is reasonable because the encoded circuits are still fault-tolerant if the phase-flip error only has $O(p\epsilon) = O(p^2)$ contribution. 

\section{Discussion}\label{sec:discussion}
We have proposed a fault-tolerant and bias-preserving gate set, namely $S$, $H$, $\mathrm{CZ}$, and $R_z$ gates that do not cause phase-flip errors at the leading order.
We have considered three use cases of our fault-tolerant computation scheme and found that the encoding is advantageous when
\begin{enumerate}
    \item the circuit is deep and consists mainly of Pauli and CNOT gates.
    \item the circuit depth is moderate, but all Clifford gates and rotation gates are allowed.
    \item the circuit is shallow, but high precision is needed.
\end{enumerate}
Another notable finding is that error correction is not necessarily mandatory. In fact, except in case 1, error correction does not reduce the error in the expectation value of physical observables.
This may be related to the inability to correct phase-flip errors with a probability of $O(p^2)$. In addition, it may also have something to do with the fact that the timing of error correction is not optimized. Exploring potential accuracy enhancements via optimization remains a prospect for future investigation.

Subsequently, we comment on the relationship of our error correction scheme to other methods that have been proposed in recent years. In the realm of fault-tolerant computation under resource constraints, error correction using flag qubits, where errors on a code of distance three can be corrected using just two ancillary qubits, is a promising approach~\cite{PhysRevLett.121.050502}. This has the advantage of requiring fewer resources than fault-tolerant syndrome measurements using a Greenberger–Horne–Zeilinger state as employed in Shor's scheme. Although we do not employ flag qubits in our scheme, the minimum number of ancillary qubits required is one as long as clean $\ket{0}$-states are supplied. If such clean states are not available, at least four ancillary qubits are needed. (See Appendix~\ref{app:state preparation}.) It is interesting whether the flag qubit method can be used to reduce this amount of resources.

Finally, we discuss the scalability of our proposed scheme. So far, we have employed a repetition code with a distance of three, but its extension to arbitrary distances remains straightforward. The repetition code with distance $d$ can correct $\lfloor (d-1)/2 \rfloor$ bit-flip errors. Using classical bounded distance decoding, the logical error rate after a single quantum operation is denoted by $p_\text{L} = 1 - \sum_{k=0}^{\lfloor (d-1)/2 \rfloor} \binom{d}{k} p^{k} (1-p)^{d-k}$. As $d$ increases, $p_\text{L}$ diminishes for most values of $p$, thereby ensuring fault-tolerant computation with arbitrarily small logical errors. However, the code rate, expressed as $1/d$, is suboptimal.

Certain logical operations necessitate postselection, raising questions about the scalability in terms of the yield of such processes. We can show that the yield after postselection does not decrease with increasing code distance as follows. Consider, for example, the logical $S$ gate. For any code distance, the logical $S$ gate is basically the same as in Fig.~\ref{fig: |+i> preparation}. We only increase the number of physical qubits by $d$. A Pauli-$Y$ error may arise solely after the action of the $S$ gate on the first qubit. Detection of such an error is feasible by measuring a single syndrome, $Z_1 Z_2$. The likelihood of detecting the $Y$ error remains constant regardless of the distance, thus the yield from postselection shows a robust insensitivity to distance variations. This characteristic is also observed for $H$, CZ, and $R_z$ gates. This property shows the scalability of our proposed scheme for arbitrary logical operations.

\section*{Acknowledgments}
We thank Suguru Endo and Yasunari Suzuki for guiding the authors to the relevant literature. We also thank Yuya O. Nakagawa for the fruitful discussion and Yasunori Lee for careful reading of and comments on the draft.

\appendix
\section{Fault-tolerant error correction}\label{app:fault-tolerant error correction}
\begin{figure}[htbp]
\begin{align*}
    \Qcircuit @C=2.em @R=2.em {
        \ustick{1}&\ctrl{3}&\ustick{8j+4}\qw&\qw     &\qw          &\qw     &\qw          &\qw     &\qw&\qw&\qw\\
        \ustick{2}&\qw     &\qw          &\ctrl{2}&\ustick{8j+5}\qw&\ctrl{3}&\ustick{8j+6}\qw&\qw     &\qw&\qw&\qw    \\
        \ustick{3}&\qw     &\qw          &\qw     &\qw          &\qw     &\qw          &\ctrl{2}&\ustick{8j+7}\qw&\qw&\qw    \\
        \ini      &\targ   &\ustick{8j+8}\qw&\targ   &\ustick{8j+9}\qw&\qw     &\qw          &\qw     &\qw&\meter \\
        \ini      &\qw     &\qw          &\qw     &\qw          &\targ   &\ustick{8j+10}\qw&\targ   &\ustick{8j+11}\qw&\meter 
        \gategroup{1}{2}{5}{10}{0.5em}{--} \\
        &&&&&\mbox{repeat $n$ times}&&&&
    }
\end{align*}
\caption{A quantum circuit of the (repeated) syndrome measurements. Numbers indicate where bit-flip errors can occur ($0 \leq j \leq n-1$).}
\label{fig: syndrome measurement}
\end{figure}
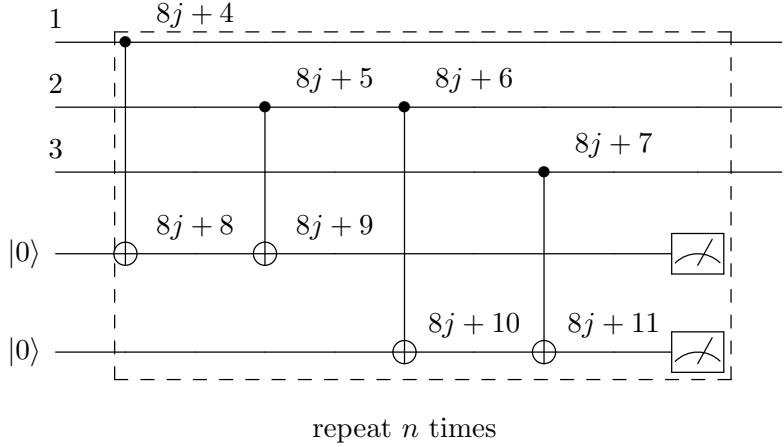

We show how to achieve fault-tolerant error correction based on the double-round syndrome measurements. 
First, we remind the reader that error correction is not fault-tolerant if the error syndrome is measured only once.
We show possible locations where bit-flip errors can arise in Fig.~\ref{fig: syndrome measurement} and error syndromes when those errors occur in Table~\ref{tab: syndrome in single round measurement}. 
Among these error patterns, the error occurring in location 5 breaks the fault tolerance of error correction. This is because the syndrome indicates that the error occurs in the third physical qubit, although the actual error occurs in the second.

\begin{table}[hbtp]
    \caption{Location of bit-flip errors and error syndromes of single-round measurement.}
    \centering
    \begin{tabular}{ccc}
    \hline
    location & $Z_1Z_2$ & $Z_2Z_3$ \\
    \hline
    1  & $-1$ & $1$  \\ 
    2  & $-1$ & $-1$  \\ 
    3  & $1$ & $-1$  \\ 
    4  & $1$ & $1$  \\ 
    5  & $1$ & $-1$  \\ 
    6  & $1$ & $1$  \\ 
    7  & $1$ & $1$ \\ 
    8  & $-1$ & $1$  \\ 
    9  & $-1$ & $1$  \\ 
    10 & $1$ & $-1$  \\ 
    11 & $1$ & $-1$  \\ 
    \hline
    \end{tabular}
    \label{tab: syndrome in single round measurement}
\end{table}

We can eliminate such an unfavorable event when we measure syndromes twice.
Table~\ref{tab: syndrome in double round measurement} lists possible error patterns and corresponding syndromes.
The bit-flip error at location 13, which is problematic in the single-round syndrome measurement, has a distinctly different syndrome than the others that trigger feedback control. Thanks to this property, fault tolerance is manifest in the error-correcting circuit with double-round syndrome measurement.
Note that the 16th and 17th patterns that are indistinguishable from the 4th pattern cause a bit-flip error on the first physical qubit, but it does not break fault tolerance.
Error correction based on the double-round measurements is not unique. For example, there is a way not to give feedback for the 4th, 16th, and 17th patterns.
\begin{table}[hbtp]
    \caption{Location of bit-flip errors and error syndromes of double-round measurement.}
    \centering
    \begin{tabular}{cccccc}
    \hline
    & \multicolumn{2}{c}{$j=0$} & \multicolumn{2}{c}{$j=1$} & \\
    location & $Z_1Z_2$ & $Z_2Z_3$ & $Z_1Z_2$ & $Z_2Z_3$ & feedback  \\
    \hline
    1  & $-1$ & $1$ & $-1$ & $1$ & $X_1$ \\ 
    2  & $-1$ & $-1$ & $-1$ & $-1$ & $X_2$ \\ 
    3  & $1$ & $-1$ & $1$ & $-1$ & $X_3$ \\ 
    4  & $1$ & $1$ & $-1$ & $1$ & $X_1$ \\ 
    5  & $1$ & $-1$ & $-1$ & $-1$ & $X_2$ \\ 
    6  & $1$ & $1$ & $-1$ & $-1$ & $X_2$ \\ 
    7  & $1$ & $1$ & $1$ & $-1$ & $I$ \\ 
    8  & $-1$ & $1$ & $1$ & $1$ & $I$ \\ 
    9  & $-1$ & $1$ & $1$ & $1$ & $I$ \\ 
    10 & $1$ & $-1$ & $1$ & $1$ & $I$ \\ 
    11 & $1$ & $-1$ & $1$ & $1$ & $I$ \\ 
    \hline
    \end{tabular}
    \begin{tabular}{cccccc}
    \hline
    & \multicolumn{2}{c}{$j=0$} & \multicolumn{2}{c}{$j=1$} & \\
    location & $Z_1Z_2$ & $Z_2Z_3$ & $Z_1Z_2$ & $Z_2Z_3$ & feedback  \\
    \hline
    12  & $1$ & $1$ & $1$ & $1$ & $I$ \\ 
    13  & $1$ & $1$ & $1$ & $-1$ & $I$ \\ 
    14  & $1$ & $1$ & $1$ & $1$ & $I$ \\ 
    15  & $1$ & $1$ & $1$ & $1$ & $I$ \\ 
    16  & $1$ & $1$ & $-1$ & $1$ & $X_1$ \\ 
    17  & $1$ & $1$ & $-1$ & $1$ & $X_1$ \\ 
    18  & $1$ & $1$ & $1$ & $-1$ & $I$ \\ 
    19  & $1$ & $1$ & $1$ & $-1$ & $I$ \\ 
    \\ 
    \\ 
    \\
    \hline
    \end{tabular}
    \label{tab: syndrome in double round measurement}
\end{table}

\section{The qubit Hamiltonian of caffeine}\label{app:caffeine hamiltonian}
We describe the details of the qubit Hamiltonian of caffeine \ce{C8H10N4O2}. The geometrical configuration of atoms is optimized using PySCF~\cite{sun2018PySCF,sun2020Recent}. 
Within the Born-Oppenheimer approximation, the second-quantized electronic Hamiltonian of caffeine is constructed by calculating molecular orbitals based on the Hartree-Fock method using the STO-3G basis set. We consider an active space of CAS(2,2) to reduce the dimension of Hilbert space. The qubit Hamiltonian is obtained via the Jordan-Wigner transformation using  OpenFermion~\cite{mcclean2020OpenFermion}. Finally, we get a tapered Hamiltonian consisting of 2 qubits specifying the eigenspace of $S_Z$, the $z$ component of the spin operator. For completeness, we show the concrete form of the tapered Hamiltonian:
\begin{align*}
    H=
    &-667.4554308557676 I 
    -0.013168856506009949 X_0 
    +0.013168856506009949 X_1 \\
    &-0.1532273887412754 Z_0 
    -0.1532273887412754 Z_1 
    +0.013169112223348517 X_0 Z_1 \\
    &-0.013169112223348517 Z_0 X_1 
    +0.025969183085931477 Z_0 Z_1 
    -0.050192647768994174 Y_0 Y_1.
\end{align*}

\section{Fault-tolerant state preparation}\label{app:state preparation}
When the physical $\ket{0}$ state cannot be prepared with high fidelity, a ``$\ket{0}$-state factory'' is needed to check for errors and make postselections.
Here, we describe a fault-tolerant way of state preparation, assuming that a bit-flip error occurs with probability $p$ even in preparing the physical $\ket{0}$ state. We assume that the noise model is identical to what we introduce in Sec.~\ref{Sec:noise model} except that clean initial states are not available.
\begin{enumerate}
    \item Prepare three physical $\ket{0}$ states.
    \item Measure $Z_1Z_2$ and $Z_2Z_3$ once each.
    \item Take out one of these states in which no bit-flip error is detected. If not, repeat this procedure until successful.
\end{enumerate}
The probability that $\ket{1}$ is still obtained incorrectly after this process is $O(p^2)$. Therefore, we can prepare $\ket{0}$ fault-tolerant by this method. Note that at least four qubits are needed in this procedure.

%

\end{document}